\documentstyle[psfig]{mn}

\newif\ifAMStwofonts

\def\lapp{\ifmmode\stackrel{<}{_{\sim}}\else$\stackrel{<}{_{\sim}}$\fi}
\def\gapp{\ifmmode\stackrel{>}{_{\sim}}\else$\stackrel{>}{_{\sim}}$\fi}
\def\psr{PSR~B1259$-$63}
\def\arcdeg{\hbox{$^\circ$}}
\def\peri{\ifmmode{\cal{T}}\else${\cal{T}}$\fi}
\newcommand{\prog}[1]{\textsc{#1}}

\title[Radio Observations of PSR~B1259--63]
{Radio Observations of PSR~B1259--63\\ through the 2004 periastron passage} 
\author[Johnston et al.]
{Simon~Johnston$^1$, Lewis~Ball$^2$, N.~Wang$^{1,3}$ \& R.~N.~Manchester$^3$\\
$^1$School of Physics, University of Sydney, NSW 2006, Australia.\\
$^2$Australia Telescope National Facility, CSIRO, P.O. Box 276, Parkes,
NSW 2870, Australia \\
$^3$Australia Telescope National Facility, CSIRO, P.O. Box 76, 
Epping, NSW 1710, Australia \\
}
\date{\today}
\pagerange{\pageref{firstpage}--\pageref{lastpage}}
\pubyear{2004}
\begin{document}
\maketitle
\label{firstpage}

\begin{abstract}
We report here on extensive radio observations
of the binary system containing \psr{} and the Be star SS 2883,
made around the time of the 2004 periastron.
This is the fourth periastron to have been observed in detail.
As in previous observations, changes in the pulsar's dispersion and rotation
measures are detected over a period spanning 200 days. We show that
the eclipse of the pulsed emission lasts from 16 days prior to
periastron to 15 days after periastron and is consistent from
one periastron to the next. We demonstrate that the timing solution
proposed by Wang et al. (2004) provides a good fit through the
2004 periastron.
The light curve of the transient unpulsed radio emission is broadly similar 
from one periastron to the next. For this periastron, however,
the light curve is strongly peaked post-periastron with rather
little enhancement prior to periastron, in contrast to the 2000
periastron where the peak flux densities were more equal.
These observations remain consistent with the interpretation
that the pulsar passes through the dense circumstellar disk
of the Be star just before and just after periastron.
The observed differences from one periastron to the next 
can be ascribed to variations in the local disk density 
and magnetic field structure at the time the pulsar enters the disk.
\end{abstract}


\section{Introduction}
\psr{} was discovered in a large-scale high frequency survey of
the Galactic plane \cite{jlm+92}. It is unique because it is
the only known radio pulsar in orbit about a massive, 
main-sequence, B2e star \cite{jml+92}. PSR B1259$-$63 has a
short spin period of $\sim$48~ms and moderate period derivative of
$2.28\times10^{-15}$, implying a 
high spin-down energy of $8.2\times 10^{35}$~erg\,s$^{-1}$ and a
characteristic age of only 330~kyr. It has a long orbital period
of $\sim$1237 days, and a large eccentricity of 0.87, with a
projected semi-major axis, $a\sin i$ of $\sim$1300~light-s.

The companion, SS 2883, is a 10th magnitude star with a mass of about
10~$M_\odot$ and a radius of 6~$R_\odot$. Assuming the pulsar
mass is 1.4~$M_\odot$ implies the
inclination of the orbit to the plane of the sky is $i\sim36\arcdeg$.
Be stars as a class have a hot, tenuous polar wind and a cooler, high density,
equatorial disk. The expected mass loss rate from a B2e star is $\sim
10^{-6}$~$M_\odot$.  Johnston et al. (1994)\nocite{jml+94} observed
H$\alpha$ emission lines from the disk at 20 stellar radii (R$_{*}$),
just inside the pulsar orbit of 24 R$_{*}$ at periastron. The density
of the disk material is high near the stellar surface
($10^{8}-10^{10}$~cm$^{-3}$), and falls off as a power-law with
distance from the star. The disk is likely to be tilted with
respect to the pulsar orbital plane, and \psr{} is eclipsed for
about 35 days as it goes behind the disk.

This pulsar has been extensively observed since its discovery 15 years ago.
Determining the timing properties of the pulsar has proved a
difficult task. Initial timing solutions obtained by Manchester et al. (1995)
and Wex et al. (1998) failed to accurately describe the data through
subsequent periastron passages. The most
recent update on the timing of \psr{} by Wang, Johnston \& Manchester (2004)
provides a comprehensive review of the many difficulties involved.
Their model shows that the pulsar glitched in late 1997 and that the timing
residuals are best modelled by including a jump in $a\sin i$ at
each periastron.
\nocite{mjl+95,wjm+98,wjm04}

Transient unpulsed radio emission seen around the time of periastron has
been discussed by Johnston et al. (1996), Johnston et al. (1999)
and Connors et al. (2002), with physical models for the emission
proposed by Melatos, Johnston \& Melrose (1995), Ball et al. (1999) and
Connors et al. (2002).
\nocite{jml+96,jmmc99,cjmm02,mjm95,bmjs99}
Unpulsed emission has also been detected in the X-ray (summarised
in Hirayama et al. 1996) and in the soft $\gamma$-ray band 
by Grove et al. (1995) and more recently Shaw et al. (2004),
with modelling of the emission described in Tavani \& Arons (1997).
\nocite{hnt+96,gtp+95,ta97,scr+04}

Kirk, Ball and Skjaeraasen (1999)\nocite{kbs99} proposed that the
system should be detectable at high energies through inverse Compton
scattering of the UV photons from the Be star by the relativistic
pulsar wind. This model has largely been confirmed through a recent
detection of the system at 
TeV energies by Aharonian et al. (2004)\nocite{aaa+04}.

\section{Observations}
\subsection{Pulsed emission}
Observations of the pulsar were carried out with the 64-m radio telescope
located in Parkes, NSW, at frequencies between 680 and 8640~MHz.

Two independent backend systems were employed; filterbank systems which
recorded total power and a correlator system which retained full
polarisation information.
The filterbank system was used at 680 and 1500~MHz only. At the lower
frequency it consists of 256 frequency channels each of width 0.250~MHz
for a total bandwidth of 64~MHz and at the higher frequency consists
of 512 frequency channels each of width 0.5~MHz for a total bandwidth
of 256~MHz. In each case the analogue signal is one-bit digitised every 
250~$\mu$s and written to tape for off-line analysis. This analysis
involved de-dispersion and folding at the topocentric period to produce
a pulse profile.
Correlator data were obtained at frequencies around 1370, 3100 
and 8640~MHz with a total
bandwidth of 256, 512 and 512~MHz at the three frequencies.
Channel bandwidths were 1~MHz and there were 512 phase bins across the
pulsar period. The data were folded on-line for an interval of 60~s
and written to disk. Off-line processing used the \textsc{Psrchive} software
application \cite{hvm04} specifically written for analysis of pulsar data.
Processing involved calibration and de-dispersion and yielded
full Stokes pulse profiles.

\begin{table}
\caption{DM and RM variations and the inferred magnetic field
for the 2004 periastron observations. The error in DM is
typically 0.2~cm$^{-3}$~pc.}
\begin{tabular}{lrcrr}
\hline & \vspace{-3mm} \\
\multicolumn{1}{c}{Date} & \multicolumn{1}{c}{Day} 
& \multicolumn{1}{c}{$\Delta$DM} & \multicolumn{1}{c}{RM}
& \multicolumn{1}{c}{$B_{\Vert}$} \\
& & (cm$^{-3}$pc) & (rad m$^{-2}$) & (mG) \\
\hline & \vspace{-3mm} \\
2004 Jan 27 & $-$39.8 & 3.0 & $-$3100 $\pm$ 300 & $-$1.3 $\pm$ 0.1 \\
2004 Jan 29 & $-$37.7 & 2.5 & $-$3800 $\pm$ 400 & $-$1.9 $\pm$ 0.2 \\
2004 Jan 31 & $-$35.7 & 3.4 & +1700 $\pm$ 200 & +0.6 $\pm$ 0.1 \\
2004 Feb 02 & $-$33.7 & 5.6 \\
2004 Feb 04 & $-$31.7 & 4.0 & $-$1500 $\pm$ 200 & $-$0.5 $\pm$ 0.1 \\
2004 Feb 06 & $-$29.7 & 3.9 \\
2004 Feb 08 & $-$27.7 & 5.9 \\
2004 Feb 10 & $-$25.7 & 4.9 & +11500 $\pm$ 1100 & +2.9 $\pm$ 0.3 \\
2004 Feb 12 & $-$23.7 & 5.3 \\
2004 Feb 14 & $-$21.7 & 5.5 \\
2004 Feb 17 & $-$18.8 & \\
2004 Feb 18 & $-$17.8 & 19.5 \\
2004 Feb 20 & $-$15.8 & \\
2004 Mar 23 & 16.1 & 3.2 \\
2004 Mar 25 & 18.1 & \\
2004 Mar 28 & 21.2 & 1.0 \\
2004 Mar 29 & 22.1 & 1.2\\
2004 Mar 31 & 24.1 & 0.8 & +6000 $\pm$ 600 & +10 $\pm$ 3 \\
2004 Apr 02 & 26.2 & 0.7 & $-$2500 $\pm$ 300 & $-$4.5 $\pm$ 1.4 \\
2004 Apr 04 & 28.2 & 0.5 & $-$280 $\pm$ 30 & $-$0.7 $\pm$ 0.3 \\
2004 Apr 08 & 32.2 & 0.4 & $-$500 $\pm$ 50 & $-$1.4 $\pm$ 0.7  \\
2004 Apr 14 & 38.2 & 0.4 & $-$1080 $\pm$ 100 & $-$3.3 $\pm$ 1.6 \\
2004 Apr 22 & 46.1 & 0.2 & $-$360 $\pm$ 40 & $-$2 $\pm$ 2 \\
\hline & \vspace{-3mm} \\
\end{tabular}
\label{dmrm}
\end{table}
In a typical observation, data were obtained at 680, 1500, 3100 and
8640~MHz in the space of 3~h. The time-of-arrival (ToA) of the fiducial
point in the pulsar profile
was calculated for each observation by convolving
with a standard template which is frequency dependent.
The templates at different frequencies were aligned
in pulse phase as shown in Figure~1 of Wang et al. (2004).
Any offset between the arrival times at the different frequencies
was then attributed to a change in dispersion measure (DM) of the pulsar
and a fit was done to obtain the offset DM.

The polarisation information allowed us to obtain the rotation measure
(RM) for each observation. There are two possible methods for
obtaining the RM. The first is to measure the change in position angle 
across the band. This can be done either at a given pulsar phase
or by time averaging across several phase bins.
The second method is to maximise the linearly polarised flux by 
choosing trial RMs. This method works well for low signal to 
noise ratios as it uses the entire pulse window.
In fact, for \psr{}, there
is very little change in the position angle across each component, 
and the two methods yield similar results.

Table~\ref{dmrm} shows the log of the observations of the pulsar
made with the Parkes telescope.
The first column gives the date of the observation and
the second column shows
the offset in days from the 2004 periastron epoch (which we denote as
\peri\ hereafter).

\subsection{Unpulsed emission}
Observations were also made with the Australia Telescope Compact Array
(ATCA), an east-west synthesis telescope located near Narrabri, NSW,
which consists of six 22-m antennas on a 6 km track.
ATCA observations can be made simultaneously at
either 1.4 and 2.4 GHz or 4.8 and 8.4 GHz with a bandwidth of 128 MHz at
each frequency subdivided into 32 spectral channels, and full Stokes
parameters.
The ATCA is also capable of splitting each correlator cycle into
bins corresponding to different phases of a pulsar's period, and
in our case the pulse period of $\sim$48~ms was split into 16 phase bins.
This allows a measurement of off-pulse and on-pulse flux densities
to be made simultaneously.

Initial data reduction and analysis were carried 
out with the \textsc{Miriad} package using
standard techniques. After flagging bad data, the primary calibrator
was used for flux density and bandpass calibration and the secondary
calibrator was used to solve for antenna gains, phases and
polarisation leakage terms. After calibration, the data
consist of 13 independent frequency channels each 8 MHz wide for each
of the 16 phase bins.
Subsequent analysis of the data was carried out as described
in Connors et al. (2002)\nocite{cjmm02}.

\section{Results}
\subsection{Pulsed emission}
\subsubsection{Timing}
\begin{figure}
\centerline{\psfig{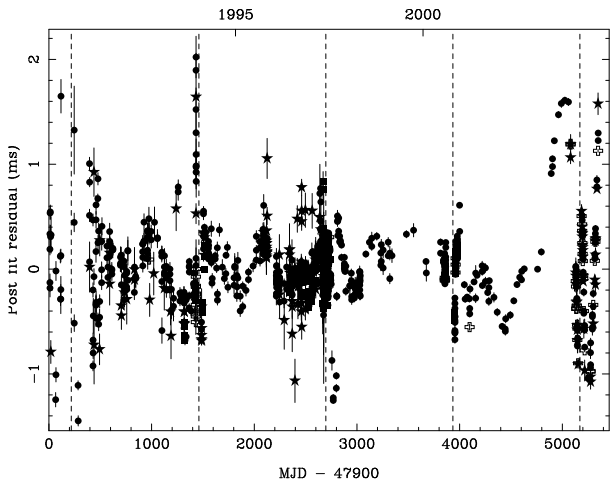}}
\caption{Post-fit timing residuals for \psr{} from 1991 January to 2004 
September with a model including jumps in $a\sin i$ at each
periastron. The rms residual is 350~$\mu$s. Dashed lines mark 
the periastron epochs.}
\label{timing}
\end{figure}

Figure~\ref{timing} shows the complete timing residuals from \psr{}.
There are more than 1200 independent timing points, with a 
data span from 1991 January to 2004 September including five
periastron passages. As described in earlier papers, the DM
of the pulsar changes near periastron and this extra DM
needs to be accounted for in the timing solution.
For this periastron, observations were made at frequencies between
0.64 and 8.4 GHz and the DM was calculated by measuring the time
offset between the TOAs at the different frequencies
(see e.g. Wex et al. 1998)\nocite{wjm+98}.

In Wang et al. (2004) we determined that the best fit solution involved
adding jumps in the value of $a\sin i$ at periastron and
that a small glitch occured at MJD 50691. 
We have now extended that model to cover the 2004 periastron passage,
obtaining successive jumps in $a\sin i$ of
60.7, $-$26.3, 2.8, 4.2 and $-$7.8~ms.
The fitted jumps for the first 4 periastrons are
within a few percent of those obtained by Wang et al. (2004). 
\begin{figure}
\centerline{\psfig{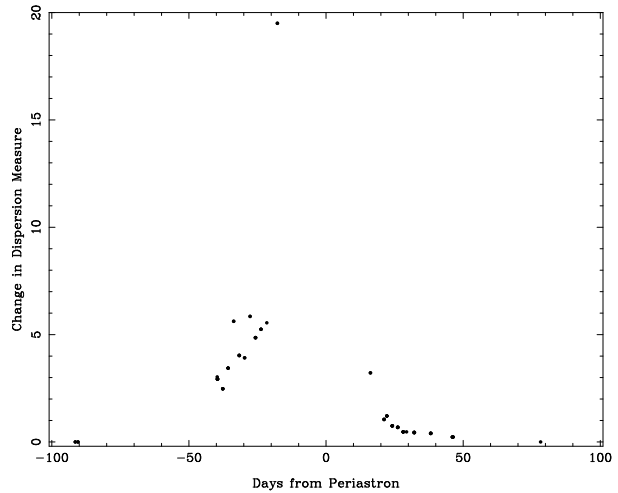}}
\caption{Dispersion Measure variations in \psr{} around the 2004
periastron. The DM contribution from the interstellar medium is assumed to be
146.7~cm$^{-3}$pc. The error bars are typically 0.2~cm$^{-3}$pc and are
smaller than the symbols.}
\label{ddm}
\end{figure}

\subsubsection{DM variations}
Column 3 of Table~\ref{dmrm} lists the DM variations and 
these are displayed in Figure~\ref{ddm}. The typical error in the DM
values are 0.2~cm$^{-3}$pc.
Although the pulsar was undetected at 1.4, 3.1 and 8.4~GHz on
\peri $-$18.8, the last detection before periastron occurred on
\peri $-$17.8 when the DM change was 19.5~cm$^{-3}$pc. Subsequent
observations on \peri $-$15.8 failed to detect the pulsar.
The eclipse lasted until \peri +16.1 although on this date
the pulsar was not detected at 1.4~GHz. Two days later, there is a
marginal detection of the pulsar at 3.1~GHz but not at higher or
lower frequencies. After the 
exit from the eclipse the DM change was only 3.2 and decayed over
an interval of $\sim$20 days. This is broadly in line
with changes seen during previous periastron passages
(see Figure~2 in Wang et al. 2004).

\subsubsection{RM variations}
\begin{table*}
   \caption{Flux densities for pulsed emission and total emission (pulsed
     plus unpulsed) for all 15 observations made with the ATCA in 2004.
     Total flux densities and errors were obtained
     from the \prog{miriad} routine \prog{imfit}.
     Pulsed flux densities and errors were obtained
     using the routine \prog{uvflux}.
     Pulsar flux densities are omitted when no significant pulses could
     be seen. The last two observations used ultra-compact arrays and the
     imaging was too poor to obtain reliable flux density estimates.}
   \begin{tabular}{|r|r|*{8}{r}|}
     \hline & \vspace{-3mm} \\
     & &  \multicolumn{8}{c}{Observing Frequency} \\
     \multicolumn{1}{c}{Date} & \multicolumn{1}{c}{Day}  & \multicolumn{2}{c}{1384 MHz} & 
     \multicolumn{2}{c}{2496 MHz} & \multicolumn{2}{c}{4800 MHz} & 
     \multicolumn{2}{c}{8400 MHz} \\
     & & \multicolumn{1}{c}{Pulsar} &
     \multicolumn{1}{c}{Total} & 
     \multicolumn{1}{c}{Pulsar} & \multicolumn{1}{c}{Total} & 
     \multicolumn{1}{c}{Pulsar} & \multicolumn{1}{c}{Total} & 
     \multicolumn{1}{c}{Pulsar} & \multicolumn{1}{c}{Total} \\
     & & \multicolumn{1}{c}{(mJy)} & \multicolumn{1}{c}{(mJy)} & 
     \multicolumn{1}{c}{(mJy)} & \multicolumn{1}{c}{(mJy)} & 
     \multicolumn{1}{c}{(mJy)} & \multicolumn{1}{c}{(mJy)} & 
     \multicolumn{1}{c}{(mJy)} & \multicolumn{1}{c}{(mJy)} \\
     \hline
Feb 14 & $-21.5$ & $2.4$\makebox[8mm][l]{$\pm 0.2$} & $4.7$\makebox[8mm][l]{$\pm 0.5$} & 
$3.3$\makebox[8mm][l]{$\pm 0.2$} & $4.8$\makebox[8mm][l]{$\pm 0.5$} & 
$2.5$\makebox[8mm][l]{$\pm 0.2$} & $2.9$\makebox[8mm][l]{$\pm 0.5$} &
$0.8$\makebox[8mm][l]{$\pm 0.2$} & $0.8$\makebox[8mm][l]{$\pm 0.3$} \\ 
Feb 21 & $-14.7$ & \multicolumn{1}{c}{---} & $12.4$\makebox[8mm][l]{$\pm 2.3$} &
\multicolumn{1}{c}{---} & $12.2$\makebox[8mm][l]{$\pm 0.5$} &
\multicolumn{1}{c}{---} & $7.8$\makebox[8mm][l]{$\pm 0.6$} &
\multicolumn{1}{c}{---} & $5.0$\makebox[8mm][l]{$\pm 0.5$} \\ 
Feb 27 & $-8.7$ & \multicolumn{1}{c}{---} & $14.7$\makebox[8mm][l]{$\pm 2.2$} &
\multicolumn{1}{c}{---} & $13.6$\makebox[8mm][l]{$\pm 0.5$} &
\multicolumn{1}{c}{---} & $10.4$\makebox[8mm][l]{$\pm 0.6$} &
\multicolumn{1}{c}{---} & $7.3$\makebox[8mm][l]{$\pm 0.7$} \\ 
Mar 04 & $-2.8$ & \multicolumn{1}{c}{---} & $16.9$\makebox[8mm][l]{$\pm 2.3$} &
\multicolumn{1}{c}{---} & $12.0$\makebox[8mm][l]{$\pm 0.6$} &
\multicolumn{1}{c}{---} & $9.9$\makebox[8mm][l]{$\pm 0.7$} &
\multicolumn{1}{c}{---} & $5.8$\makebox[8mm][l]{$\pm 0.8$} \\ 
Mar 10 & $+3.2$ & \multicolumn{1}{c}{---} & $13.2$\makebox[8mm][l]{$\pm 1.8$} &
\multicolumn{1}{c}{---} & $14.2$\makebox[8mm][l]{$\pm 0.5$} &
\multicolumn{1}{c}{---} & $10.5$\makebox[8mm][l]{$\pm 0.8$} &
\multicolumn{1}{c}{---} & $7.6$\makebox[8mm][l]{$\pm 0.6$} \\ 
Mar 15 & $+8.2$ & \multicolumn{1}{c}{---} & $12.5$\makebox[8mm][l]{$\pm 0.9$} &
\multicolumn{1}{c}{---} & $9.8$\makebox[8mm][l]{$\pm 0.7$} &
\multicolumn{1}{c}{---} & $4.8$\makebox[8mm][l]{$\pm 1.4$} &
\multicolumn{1}{c}{---} & $1.5$\makebox[8mm][l]{$\pm 0.7$} \\ 
Mar 23 & $+16.1$ & \multicolumn{1}{c}{---} & $20.0$\makebox[8mm][l]{$\pm 1.0$} & 
$0.6$\makebox[8mm][l]{$\pm 0.2$} & $17.8$\makebox[8mm][l]{$\pm 0.5$} & 
$0.6$\makebox[8mm][l]{$\pm 0.2$} & $10.2$\makebox[8mm][l]{$\pm 1.8$} &
$0.2$\makebox[8mm][l]{$\pm 0.2$} & $4.3$\makebox[8mm][l]{$\pm 1.0$} \\ 
Mar 28 & $+21.1$ & $3.2$\makebox[8mm][l]{$\pm 0.2$} & $51.2$\makebox[8mm][l]{$\pm 2.9$} & 
$3.6$\makebox[8mm][l]{$\pm 0.2$} & $44.3$\makebox[8mm][l]{$\pm 2.0$} & 
$2.4$\makebox[8mm][l]{$\pm 0.2$} & $27.5$\makebox[8mm][l]{$\pm 2.9$} &
$1.9$\makebox[8mm][l]{$\pm 0.2$} & $14.9$\makebox[8mm][l]{$\pm 3.2$} \\ 
Apr 04 & $+28.1$ & $5.1$\makebox[8mm][l]{$\pm 0.2$} & $27.0$\makebox[8mm][l]{$\pm 1.4$} & 
$3.6$\makebox[8mm][l]{$\pm 0.2$} & $22.0$\makebox[8mm][l]{$\pm 0.7$} & 
$1.6$\makebox[8mm][l]{$\pm 0.2$} & $13.0$\makebox[8mm][l]{$\pm 1.1$} &
$1.1$\makebox[8mm][l]{$\pm 0.2$} & $7.5$\makebox[8mm][l]{$\pm 1.2$} \\ 
Apr 09 & $+33.5$ & $4.6$\makebox[8mm][l]{$\pm 0.2$} & $27.6$\makebox[8mm][l]{$\pm 3.6$} & 
$3.1$\makebox[8mm][l]{$\pm 0.2$} & $ 9.7$\makebox[8mm][l]{$\pm 2.0$} & 
$1.5$\makebox[8mm][l]{$\pm 0.2$} & $8.5$\makebox[8mm][l]{$\pm 2.0$} &
$0.8$\makebox[8mm][l]{$\pm 0.2$} & $4.0$\makebox[8mm][l]{$\pm 1.5$} \\ 
Apr 14 & $+38.5$ & $5.9$\makebox[8mm][l]{$\pm 0.2$} & $20.5$\makebox[8mm][l]{$\pm 3.8$} & 
$5.7$\makebox[8mm][l]{$\pm 0.2$} & $19.4$\makebox[8mm][l]{$\pm 2.6$} & 
$2.7$\makebox[8mm][l]{$\pm 0.2$} & $7.5$\makebox[8mm][l]{$\pm 1.8$} &
$1.8$\makebox[8mm][l]{$\pm 0.2$} & $3.8$\makebox[8mm][l]{$\pm 0.7$} \\ 
May 10 & $+64.2$ & $2.2$\makebox[8mm][l]{$\pm 0.2$} & $10.4$\makebox[8mm][l]{$\pm 0.7$} & 
$2.5$\makebox[8mm][l]{$\pm 0.2$} & $10.7$\makebox[8mm][l]{$\pm 0.5$} & 
$1.1$\makebox[8mm][l]{$\pm 0.2$} & $5.8$\makebox[8mm][l]{$\pm 0.9$} &
$0.5$\makebox[8mm][l]{$\pm 0.2$} & $3.7$\makebox[8mm][l]{$\pm 0.7$} \\ 
Jun 09 & $+94.0$ & $3.2$\makebox[8mm][l]{$\pm 0.2$} & $11.1$\makebox[8mm][l]{$\pm 1.8$} & 
$2.8$\makebox[8mm][l]{$\pm 0.2$} & $4.0$\makebox[8mm][l]{$\pm 0.6$} & 
$1.4$\makebox[8mm][l]{$\pm 0.2$} & $1.3$\makebox[8mm][l]{$\pm 0.3$} &
$0.6$\makebox[8mm][l]{$\pm 0.2$} & $0.2$\makebox[8mm][l]{$\pm 0.2$} \\ 
Aug 04 & $+150.1$ & $6.7$\makebox[8mm][l]{$\pm 0.2$} & \multicolumn{1}{c}{---} & 
$4.6$\makebox[8mm][l]{$\pm 0.2$} &\multicolumn{1}{c}{---} & 
$2.2$\makebox[8mm][l]{$\pm 0.2$} &\multicolumn{1}{c}{---} &
$0.3$\makebox[8mm][l]{$\pm 0.2$} &\multicolumn{1}{c}{---} \\ 
Sep 09 & $+186.1$ & $4.8$\makebox[8mm][l]{$\pm 0.2$} &\multicolumn{1}{c}{---} & 
$2.8$\makebox[8mm][l]{$\pm 0.2$} &\multicolumn{1}{c}{---} & 
$2.4$\makebox[8mm][l]{$\pm 0.2$} &\multicolumn{1}{c}{---} &
$1.6$\makebox[8mm][l]{$\pm 0.2$} &\multicolumn{1}{c}{---} \\ 
        \hline
   \end{tabular}
   \label{flux}
\end{table*}
Column 4 of Table~\ref{dmrm} shows the RM as a function of epoch.
The typical error bars are of order 10 per cent.
The RM changes significantly both in magnitude and in sign between
the observations but there is little evidence of a change in RM
within the duration of a single observation.
On Feb 2 (\peri $-$34) we were unable to measure an RM even at 8.4~GHz,
on \peri $-$30 the polarisation quality of the data is poor 
and on \peri $-$28 the pulsar
is very weak and no RM information could be extracted. Following
the very large RM value on \peri $-$26, the pulsar appeared completely
depolarised at all frequencies. After the pulsar re-emerged from
the eclipse, no RMs could initially be measured. A high, positive
RM was measured on \peri +24, subsequent values were negative thereafter.
It is noticeable that the RM varies by more than
a factor of 10 in the post-periastron observrations and yet the 
DM varies only by a factor of 2.
There are three potential explanations for the depolarisation of the signal.
It is possible that the RM is so high that even across a single frequency
channel the position angle varies significantly, thus depolarising
the signal. This would imply an RM greater than $10^5$~rad~m$^{-2}$
for the 8.4~GHz observations. Secondly, the RM could be highly variable
on the timescale of a few minutes, causing depolarisation when
time averaging occurs. Finally, there may be different values of RM
along different ray paths, especially when the pulsar becomes scatter
broadened. We believe it is likely that all three occur close to
the point of eclipse of the pulsar.

The RM values measured are significantly different from those obtained
by Connors et al. (2002) for the 2000 periastron.
In 2000 the pulsar was depolarised from \peri $-$46 onwards,
whereas in 2004 polarisation was detected up to \peri $-$27.
Large and variable polarisation was detected following periastron
in both 2000 and 2004, although the large value of $-$7700~rad~m$^{-2}$
measured in 2000 was not repeated in 2004.

Armed with both the DM and the RM we can compute the 
magnetic field parallel to the line of sight, $B_{\Vert}$ in mG via
\begin{equation}
B_{\Vert} \,\,\, = 1.232\times 10^{-3} \,\,\, \frac{{\rm RM}}{{\rm \Delta DM}}
\end{equation}
Values of $B_{\Vert}$ are shown in column 5 of Table~\ref{dmrm}. The
error bar in $B_{\Vert}$ is dependent on the RM error for the
pre-periastron data and the DM error for the post-periastron data.

\subsection{Unpulsed emission}
Table~\ref{flux} lists the pulsed flux density and the total
flux density detected at the four frequencies observed with the ATCA.
Dashes indicate that the pulsar was not detected at that epoch.
The dates of these observations were carefully selected to provide
good sampling of the transient unpulsed emission based on the
observations of the previous three periastron passages.
The non detection of the pulsar at \peri $-$14.7 is consistent with 
the lack of pulsed emission observed at Parkes one day earlier.
The re-appearance of the pulsar at \peri +16.1 at the higher frequencies
is also consistent with the Parkes observations.

\begin{figure}
\centerline{\psfig{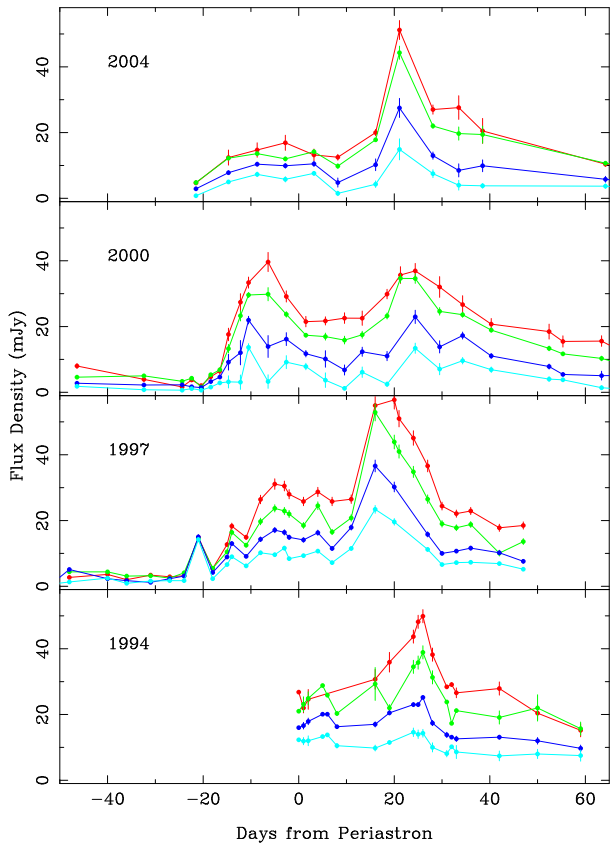}}
\caption{Transient radio emission around the time of periastron for
four different periastron passages. Each plot displays four different
frequencies, 1384, 2496, 4800 and 8640~MHz,
in order from top down. The 1994 and 1997 data
were previously published in Johnston et al. (1999) and the 2000 data
published by Connors et al. (2002).}
\nocite{jmmc99,cjmm02}
\label{fluxfig}
\end{figure}

Figure~\ref{fluxfig} shows the flux density of the transient unpulsed 
emission detected from the \psr{} system through the 2004 periastron passage
(top panel), together with the corresponding data from the previous
three periastron passages (lower panels).
Although the sampling rate was not as high in 2004 as in either
1997 or 2000, it is unlikely that any significant feature of the
light curve has been missed.

Unpulsed emission was detected at a low level on the first day that
observations were made, at \peri $-$22.
The flux density at all four frequencies observed then increased
steadily, reaching a maximum ($\sim$17~mJy at 1.4~GHz) between \peri $-$10 
and \peri $-$3, and subsequently decreased to a minimum around \peri +7.
The flux density at all four frequencies then increased quite rapidly
peaking around \peri +22 (nearly 50~mJy at 1.4~GHz), significantly
higher than the level attained prior to periastron.
Just 5 days later, at \peri +27, the flux density had dropped to
around half its peak value, and it subsequently decreased more slowly.
The unpulsed emission was still detectable at levels of a few mJy
at the time of the final observation around \peri +65.
There is some suggestion of absorption at frequencies below 2.4~GHz
in the emission observed at the earliest times, \peri $-$22 and
\peri $-$15.
At other epochs the spectrum is well fitted by a simple power law of
the form $S_{\nu}= C\nu^{\alpha}$ with $\alpha \sim -0.6$.

\section{Comparison of Periastron Passages}
\begin{figure}
\centerline{\psfig{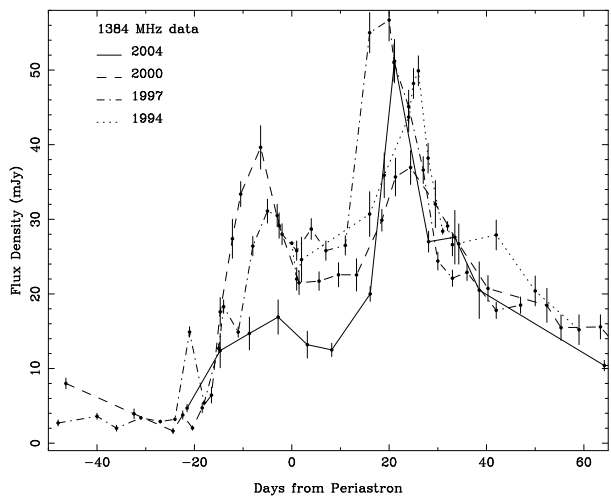}}
\caption{Light curves of the unpulsed emission at 1384~MHz from the 
four periastron passages. Solid line denotes 2004 data, dashed line
denotes 2000 data, dash-dot line is 1997 data and dotted line is
1994 data.}
\label{20cm}
\end{figure}

The top panel of Figure~\ref{fluxfig} shows the light curve for
each of the four frequencies from the 2004 periastron.
Below that panel we show the data from the three preceding periastron
passages in 1994, 1997 and 2000.
Figure~\ref{20cm} shows the unpulsed 
emission at 1384~MHz from all four periastron passages superposed,
to allow for easy comparison of the observed light curves.

The similarities between the four periastron passages are striking.
In the three cases where pre-periastron observations were obtained,
the unpulsed emission was detectable by \peri $-$20,
subsequently increased to a maximum around \peri $-$7 and then decayed
slowly to a local minimum around \peri +10.
In all four cases the flux density after periastron increased from a
local minimum around \peri +10 to a maximum close to \peri +20, then
decreased relatively rapidly until \peri +30, and subsequently declined
more slowly.

The general behaviour of the unpulsed emission has been interpreted
as resulting from the interaction of the pulsar and its wind with
the dense disk surrounding the Be star companion
\cite{bmjs99}.
The pulsar passes through the disk twice, once a few days before
periastron and then again a few days afterward.
The unpulsed emission has been broadly interpreted as resulting from
two emission regions associated with these interactions.
                                                                                
However, there are also significant differences between the light
curves at the four periastron passages which appear to be robust
despite the different data sampling intervals.
In 1997 and 2000 the flux density rises rapidly to its pre-periastron 
value over $\sim 10$~days.
In 2004 the increase apparently occurs more slowly,
at least at 1.4 and 2.4~GHz, but this could be due to decreasing
optical depth.
In 1997 and 2004 the first maximum is not well defined, but in 2000
it appears as a distinct peak between \peri $-$10 and \peri $-$8.
In 1997 and 2004 the peak flux detected after periastron is approximately
twice that seen before periastron, but in 2000 the peak before
periastron is comparable to that after periastron.
In 1994, 2000 and 2004 the peak after peristron occurs between 20 and
25 days after periastron, while in 1997 the post-peristron peak is
significantly earlier, occurring between \peri +15 and \peri +20.
                                                                                
These differences in the unpulsed emission most likely reflect
differences in the properties of the disk at the site of the
pulsar crossings.
The behaviour of the pulsed emission provides a coarse probe of the
disk geometry, since the disappearance of pulsed emission
pre-periastron, and its reappearance post-periastron, are interpreted
as marking the times when the pulsar passes into the disk and then
re-emerges from behind it.
The epochs of the observations which bracket the start and end of the
eclipse of the pulsed emission are summarised in Table~\ref{eclipse}.
                                                                                
\begin{table}
\caption{Observations bracketting the disappearance and reappearance
of pulsed emission at the four periastron epochs.
Asterisks indicate days on which the source was both detected and not
detected in separate observations.}
\begin{tabular}{lcccr@{.}l}
\hline & \vspace{-3mm} \\
\multicolumn{1}{c}{periastron} & \multicolumn{4}{c}{day number}\\
\multicolumn{1}{c}{epoch} & \multicolumn{1}{c}{last} & \multicolumn{1}{c}{first non-} & 
    \multicolumn{1}{c}{last non-} & \multicolumn{2}{c}{first} \\
& \multicolumn{1}{c}{detection} & \multicolumn{1}{c}{detection} &
    \multicolumn{1}{c}{detection} & \multicolumn{2}{c}{detection} \\
\hline & \vspace{-3mm} \\
1994 Jan 9  & $-$20.1 & $-$     & 13.9 & ~~~23&9 \\
1997 May 29 & $-$18.9 & $-$17.5 & 13.2 & 15&2* \\
2000 Oct 17 & $-$18.4 & $-$16.5 & 13.3 & 18&5 \\
2004 Mar 7  & $-$18.3 & $-$15.3 & 13.7 & 15&7* \\
\hline \\
\end{tabular}
\label{eclipse}
\end{table}

The best constrained are the epochs of re-appearance after peristron in
1997 and 2004.
On both these occasions observations with the Parkes telescope
on \peri +15 included
both detections and non-detections of the pulsed emission --- consistent
with that being the actual day of re-emergence.
The epoch of the pulsar disappearance before periastron in 1997 is
also well constrained.
Parkes observations on \peri $-$18.9 detected the pulsar and ATCA
observations on \peri $-$17.5 did not. 
Unfortunately the other eclipse epochs are less well determined, but
all are consistent with an eclipse starting on \peri $-$18 and ending
between \peri +14 and \peri +15.

Comparison of the epochs of the pulsar eclipse 
with the light curve of the unpulsed
emission suggests the following.
Just before the pulsar goes into the eclipse prior to periastron, and 
at the time when the dispersion measure has changed dramatically,
the unpulsed emission starts to rise.
The implications are that the onset of the unpulsed emission occurs as
the pulsar enters the Be star disk.
The re-detection of the pulsed emission occurs just before the
post-periastron peak of the unpulsed emission.
This is consistent with the peak of the unpulsed emission coinciding
with the end of the interaction between the pulsar and the disk, and
hence the ceasing of the supply of relativistic electrons to the
pulsar wind -- Be star disk bubble.

\begin{figure}
\centerline{\psfig{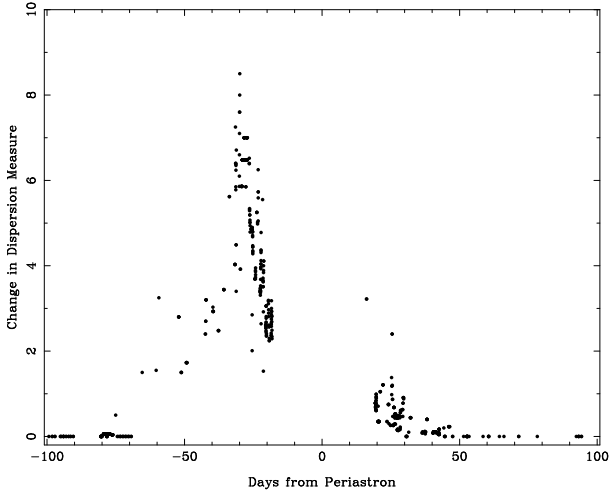}}
\caption{DM variations as a function of days from periastron
after combining all data from 1990 January to 2004 September.
The error bars are smaller than the symbols.
The DM contribution from the interstellar medium is assumed to be
146.7~cm$^{-3}$pc.}
\label{bigdm}
\end{figure}

Figure~\ref{bigdm} combines all the DM variations measured since 1990
onto a single plot. Again, the results are largely consistent from
one periastron to the next (although the sampling is far from uniform).
The eclipse of the pulsed emission is clearly delineated.
The asymmetry between the pre- and post-periastron data is striking.
This is largely because of the geometry of the system; the
pulsar is on the far side of the Be star
with respect to the observer prior to periastron and on the near
side following periastron. The path length is therefore correspondingly 
larger resulting in a larger DM variations pre-periastron.

\section{Discussion}
The differences between the light curves of the unpulsed emission
observed through the 1997, 2000 and 2004 periastron passages are
difficult to reconcile with the simple models so far proposed.
Melatos, Johnston \& Melrose (1995)\nocite{mjm95}
established that free-free absorption
in an inclined disk provided a good fit
to the observed variations in DM and RM through periastron.
                                                                                
Ball et al.\ (1999)\nocite{bmjs99} proposed that the unpulsed emission was
synchrotron radiation from two populations of electrons impulsively
accelerated as the pulsar passed through the Be star disk before and after
periastron.
They showed that the qualitative behaviour of the unpulsed emission
was consistent with a situation where two such sources cooled
primarily as a result of synchrotron losses.
The rise times to the two emission peaks were taken to be the same,
around 10 days, corresponding the the interaction time between the
pulsar wind and the Be star disk, roughly twice the time taken 
for the pulsar itself to pass through the disk. 
The unpulsed emission observed during the 2000 periastron showed a
somewhat different behaviour,
and Connors et al.\ (2002)\nocite{cjmm02} presented a
model based on that from Ball et al.\ (1999)\nocite{bmjs99},
but with the two synchrotron emitting regions
cooling primarily through adiabatic losses as the emitting region is
advected outward in the expanding radial flow of the Be star disk.

The observed epochs of the eclipse of the pulsed emission provide
little evidence of gross differences in the disk geometry between
the different periastron passages, 
although the sparse sampling of the observations does not rule out some
variation between the periastron eclipse epochs.
On the other hand, differences in the DM and RM variations between
periastron passages as the pulsar enters and re-emerges from eclipse
indicate that there are significant differences in the local properties
of the Be star wind/disk encountered by the pulsar.
These local properties, particularly the number density of the wind
and disk, will have a direct effect on determining the properties of
the pulsar wind bubble and hence the observed radio emission from the 
pre-periastron and post-periastron encounter.
                                                                                
If synchrotron losses dominate, the initial flux density decay 
following each peak is linear, with the decay time of the pre- and 
post-periastron sources determined by the magnetic field strength 
within the relevant source region.
Ball et al.\ (1999)\nocite{bmjs99} proposed that the 1997 observations were
consistent with a decay time of approximately 72 days for emission
region associated with the pre-periastron disk crossing, and just 9
days for the post-periastron source.
Taking \peri $-$18 and \peri +13 as the epochs of the pre- and
post-periastron crossings, the corresponding
orbital separations are $40~R_{*}$ and $33~R_{*}$.
If the disk of the Be star was homogeneous this would imply
a stronger magnetic field -- and hence shorter decay time
-- for the post-periastron source than the pre-periastron source,
because the smaller pulsar/Be star separation
will result in a smaller pulsar wind bubble.
After the initial linear decay, synchrotron losses produce a
distinct break in the light curve once losses become significant
at the energy of the electrons responsible for the emission at
the observing frequency.
This break is followed by a rapid frequency-dependent decay
in the radio flux.
No obvious evidence for such a break has been detected.
                                                                                
Connors et al.\ (2002)\nocite{cjmm02} suggested that the
unpulsed emission associated with the 2000 periastron fitted better
with adiabatic losses as the dominant mechanism for the decay.
In this case the flux density decay follows a power law, with the
difference between the decay of the pre- and post-periastron sources
determined by the relevant pulsar Be star separation.
This provides a good qualitative match to the 2000 periastron light
curves although it is difficult to account
for the emission plateau between \peri\ and \peri +15.
This model does not provide a good fit to the emission observed in 1997.

The 2004 light curves are difficult to fit with either of these models
for two reasons.
The first phase of emission prior to \peri +5, with its apparently
slower increase and poorly-defined peak, is qualitatively different to
that seen in either 1997 or 2000.
The combination of a rapid initial post-periastron decrease
followed by a much slower decrease is similar to the behaviour in
1997 and 1994, but at the time of the transition the flux is 
much higher than the extrapolated pre-periastron decrease.

The system has recently been detected at TeV energies by
Aharonian et al. (2004)\nocite{aaa+04} in observations
using the HESS telescopes.
An initial detection was made
10 days prior to the 2004 periastron. After a gap in the observing because
of the full moon, observations started again
at \peri +10, after an initial non-detection the flux
rose sharply, peaking at \peri +20. It then began a slow
decay with some detectable flux still present 100 days after periastron.
The maximum flux above 230~GeV was 7 per cent of that of the Crab
Nebula, and the spectrum between 0.4 and 3~TeV is well fitted by a
power law with photon index $\sim$2.7.
Both the amplitude and the spectrum were close
to that predicted by the Kirk et al. (1999)\nocite{kbs99} model in which
the Be star photons are scattered by electrons in the shocked region
of the pulsar wind.
The observed TeV light curve is poorly sampled, but appears remarkably
similar to the radio light curve.
Aharonian et al. (2004) \nocite{aaa+04} show that
the TeV peak following periastron occurs at approximately the same epoch
as the second radio peak, and like the radio flux the TeV gamma
ray emission decays over 100 days or so.

\section{Conclusions}
It is now apparent that while the gross features of the radio emission
from the \psr{}/SS2883 system are repeated through each periastron
passage, the details of both the pulsed and the transient unpulsed
emission differ considerably from one periastron to the next.

The data from the four periastron encounters so far observed
at radio frequencies remain consistent with the interpretation that
the pulsar passes through a dense disk around its Be star companion
just before peristron and is obscured for some 33 days before
re-emerging.
The data imply that there is little difference between the
geometry of the encounter between the pulsar and the Be star disk from one
pass to the next.
Unpulsed emission appears near the onset of the encounter between the 
pulsar and the Be star disk and the proposal that it originates 
from the two regions where the pulsar passes through the disk
remains consistent with the observations.
However, there are significant differences between the DM and
RM of the pulsar and the transient unpulsed emission
from one periastron to the next.
These differences all suggest that the local
properties of the Be star disk encountered by the pulsar,
such as the number density and magnetic field,
vary considerably between periastron passages.
Modelling the detailed dependence of emission observed through each
individual periastron is therefore unlikely to be productive.

The similarity between the light curves of the unpulsed radio emission
and the TeV emission is intriguing.
It raises an apparent contradiction, in that the radio emission is
most likely associated with the pulsar--disk interaction while the
properties of the TeV emission are very well fitted by model
predictions that do not involve the disk.
This is certain to become clearer as more detailed comparisons of
the radio, X-ray and gamma-ray observations and the models of the system
are completed.

\section*{Acknowledgments}
The Australia Telescope is funded by the Commonwealth of 
Australia for operation as a National Facility managed by the CSIRO.
We are grateful to Willem van Straten and Aidan Hotan for constant
software support throughout this project.

\bibliography{modrefs,psrrefs,crossrefs}
\bibliographystyle{mn}
\label{lastpage}
\end{document}